\RequirePackage{fix-cm}
\documentclass[smallextended]{svjour3}       
\smartqed  
\newcommand{\Slash}[1]{{\ooalign{\hfil/\hfil\crcr$#1$}}}
\newcommand{\tr}{{\rm tr}}

\usepackage{graphicx}
\usepackage{amsmath}	
\begin{document}

\title{The $\eta'N$ interaction from a chiral effective model and
$\eta'$-$N$ bound state\footnote{Reprot No.: KUNS-2530}
}


\author{Shuntaro Sakai         \and
        Daisuke Jido 
}


\institute{S.~Sakai \at
              Department of Physics, Graduate School of Science, Kyoto
              University, Kyoto 606-8502, Japan \\
              Tel.: +81-75-753-3820\\
              Fax: +81-75-753-3838\\
              \email{s.sakai@ruby.scphys.kyooto-u.ac.jp}           
           \and
           D.~Jido \at
              Department of Physics Tokyo Metropolitan University
           Hachioji, Tokyo 192-0397, Japan
}

\date{Received: date / Accepted: date}

\maketitle

\begin{abstract}
The $\eta'$ mass reduction in the nuclear medium is expected from the
 degeneracy of the pseudoscalar-singlet and octet mesons when chiral
 symmetry is manifest.
In this study, we investigate the $\eta'N$ 2body interaction which is
 the foundation of the in-medium $\eta'$ properties using the linear
 sigma model as a chiral effective model.
The $\eta'N$ interaction in the linear sigma model comes from the scalar
 meson exchange with U$_A$(1) symmetry effect and  is found to
 be fairly strong attraction.
Moreover, the $\eta N$ transition is included in our calculation, and is
 important for the imaginary part of the $\eta'$-optical potential.
The transition amplitude of $\eta'N$ to the $\eta N$ channel is relatively small
 compared to that of elastic channel.
From the analysis of the $\eta'N$ 2body system, we find a $\eta'N$ bound
 state with the binding energy $12.3-3.3i$MeV.
We expect that this strongly attractive two body interaction leads to a deep and
 attractive optical potential.
\end{abstract}
\newpage
\section{Introduction}
\label{intro}
The study of the meson properties in nuclear medium is one of the most
exciting topics in the hadron physics.
The meson property is strongly related to the non-perturbative natures of
Quantum Chromodynamics (QCD). 
Especially the $\eta'$ meson has a strong connection to the U$_A$(1)
anomaly and the spontaneous breaking of chiral symmetry.
In the ordinary explanation, the large mass of the $\eta'$
meson should be attributed to the explicit breaking of the 
U$_A$(1) symmetry due to the anomaly.
However, the chiral symmetry breaking is also responsible for the
generation of the $\eta'$ mass \cite{Cohen1996,Lee1996,Evans1996,Jido2012}.
In Refs.~\cite{Cohen1996,Lee1996,Evans1996,Jido2012}, they have argued the
degeneracy of the pseudoscalar singlet and
octet mesons when chiral symmetry is fully restored in three flavor
system independently of the U$_A$(1) symmetry.
Thus, both the U$_A$(1) anomaly and the chiral symmetry breaking
are essential for the generation of the $\eta'$ mass.
The effective restoration of the
U$_A$(1) symmetry is also pointed out from the viewpoint of the instanton
dynamics \cite{Shuryak1982}.
The in-medium properties of the $\eta'$ meson concerned with the
in-medium U$_A$(1) symmetry is interested in for a
long time (see, for example, Refs.~\cite{Pisarski1984,Kunihiro1989}).

One of the recent interest of the in-medium $\eta'$ property is
related with the partial restoration of chiral symmetry in the
nuclear medium \cite{Jido2012}.
The partial restoration of chiral symmetry means the reduction of the
quark condensate in low density systems.
The in-medium quark condensate
$\left<\bar{q}q\right>_\rho$ is given as
\begin{align}
 \left<\bar{q}q\right>_\rho=\left<\bar{q}q\right>_{\rho=0}\left(1-\frac{\sigma_{\pi
 N}}{f_\pi^2m_\pi^2}\rho\right),
\end{align}
in small density systems \cite{Drukarev1991}.
Some experiments of the pion-nucleus system suggest
about 35\% reduction of the quark condensate at the normal nuclear
density \cite{Suzuki2004,Friedman2004}.
Assuming small change of the $\eta$ meson mass in the nuclear
medium, we expect large mass reduction of the $\eta'$ mass in the
nuclear medium.
In the finite temperature and density system, there are some model calculations
focusing on the $\eta'$ mass
\cite{Kunihiro1989,Bernard1988,Costa2003,Nagahiro2006,Bass2006,Sakai2013}.
Furthermore, the interpretation of the mass reduction suggests possible
$\eta'$-mesonic nuclei as the attractive optical potential in the
nuclear medium \cite{Nagahiro2005}.

There are some analyses of the experimental data related to the
$\eta'$ properties.
From the analysis of the transparency ratio, the optical potential of
the $\eta'$ meson is estimated \cite{Nanova2013}.
The $\eta'N$ scattering length is extracted from the $pp\rightarrow
pp\eta'$ reaction \cite{Moskal2000,Moskal2014}.

The purpose of our work is to study the $\eta'$-optical potential.
In this study, we analyze the $\eta'N$ system with the linear sigma
 model.
Advantages to use the linear sigma model in this study are as follows;
it shares common symmetry feature to QCD, such as the three flavor
chiral symmetry and the U$_A$(1) anomaly.
It is easy to demonstrate the spontaneous chiral symmetry breaking and
its partial restoration in the nuclear medium.
Here we assume that the chiral symmetry is partially restored at the
normal nuclear density with 35\% reduction of the quark condensate.
The nucleon degree of freedom can be implemented into the linear sigma
model as a fundamental field.

In the following sections, we explain the model setup and show the
results of our analysis of the $\eta'N$ system.

\section{The linear sigma model and the $\eta'N$
 interaction\label{lsm}}
In this section, we explain the setup to evaluate the $\eta'N$
interaction \cite{Sakai2013}.

The Lagrangian used in our calculation given as follows;
\begin{align}
  \mathcal{L}=&\frac{1}{2}\tr \partial_\mu M\partial^\mu
 M^\dagger-\frac{\mu^2}{2}\tr MM^\dagger
 -\frac{\lambda}{4}\tr(MM^\dagger)^2-\frac{\lambda'}{4}\left[\tr(MM^\dagger)\right]^2\notag\\
&+A\tr(\chi
 M^\dagger+\chi^\dagger M)+\sqrt{3}B(\det M+\det M^\dagger) \notag\\
 &+\bar{N}(i\Slash{\partial}-m_N)N-\bar{N}g\left(\frac{\tilde{\sigma_0}}{\sqrt{3}}{\bf
 1}+\frac{\tilde{\sigma_8}}{\sqrt{6}}{\bf
 1}\right)N\notag\\
&-\bar{N}ig\gamma_5\left(\frac{\vec{\pi}\cdot\vec{\tau}}{\sqrt{2}}+\frac{\eta_0}{\sqrt{3}}{\bf 1}+\frac{\eta_8}{\sqrt{6}}{\bf 1}\right)N,\label{lag}
\end{align} 
where $\tau_a$ and $\lambda_a$ are the Pauli and Gell-Mann matrices
normalized as $\tr \lambda_a\lambda_b=\tr\tau_a\tau_b=2\delta_{ab}$, respectively, and
\begin{align}
 M=&\sum_{a=0}^8\frac{\sigma_a\lambda_a}{\sqrt{2}}+i\sum_{a=0}^8\frac{\pi_a\lambda_a}{\sqrt{2}},\\
N=&
\begin{pmatrix}
 p\\
 n
\end{pmatrix},\ \chi=\sqrt{3}
\begin{pmatrix}
 m_u&&\\
 &m_d & \\
 & &m_s 
\end{pmatrix}=
\sqrt{3}\begin{pmatrix}
 m_q&&\\
 &m_q & \\
 & &m_s 
\end{pmatrix},\label{mq}\\
  {\tilde \sigma_0}=&\sigma_0-\left<\sigma_0\right>,\ {\tilde
 \sigma_8}=\sigma_8-\left<\sigma_0\right>,\  m_N=\frac{g}{\sqrt{3}}\left(\left<\sigma_0\right>+\frac{\left<\sigma_8\right>}{\sqrt{2}}\right).\label{mnlsm}
\end{align}
This Lagrangian is constructed to be invariant under the chiral
transformation of hadron fields with the explicit breaking due to the
quark mass.
The baryon field is assumed to belong to the $(\bar{\bf 3}, {\bf
3})\oplus({\bf 3},\bar{\bf 3})$ representation of the SU(3)$_L \otimes$
SU(3)$_R$ group and here we show only the relevant nucleon field and its
coupling to the non-strange mesons.
In Eq.~(\ref{mq}), we assume the isospin symmetry with $m_q=m_u=m_d$, and
introduce the explicit flavor symmetry breaking with $m_q\neq m_s$.
Due to the flavor symmetry breaking, the observed $\eta$ and $\eta'$ are
linear combinations of the SU(3) eigenstate, $\eta_0$ and $\eta_8$.
The free parameters contained in the Lagrangian,
$\mu^2,\lambda,\lambda',A,B,m_q,m_s$, are determined to reproduce the
observed meson masses and decay constants and the 35\% reduction of the
quark condensate at the normal nuclear density.
The nuclear medium is introduced with the nucleon mean-field
approximation.

\begin{figure}[t]
 \centering
 \includegraphics[width=7cm]{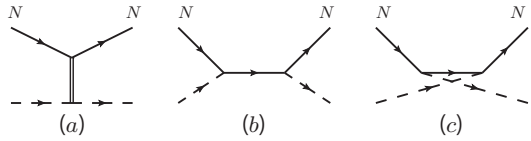}
 \caption{The diagrams contributing to the $\eta'N$ scattering
 amplitude.
The dashed, solid, and double-solid lines represent the pseudoscalar,
 nucleon, and scalar meson propagation.}
 \label{epN_diag}
\end{figure}

From this Lagrangian, we can obtain the $\eta'N$ scattering amplitude.
At the tree level, the diagrams contributing to the scattering amplitude are
shown in Fig.~\ref{epN_diag}.
The diagram ($a$) is the contribution from the scalar meson exchange,
and the diagrams ($b$) and ($c$) are the Born terms.

Especially in the chiral limit, the $\eta'N$ scattering amplitude
$V_{\eta'N}$, and $\eta'N$ transition amplitude to $\eta N$ channel
$V_{\eta'N\rightarrow\eta N}$ are given as
\begin{align}
 &V_{\eta'N}=-\frac{6gB}{\sqrt{3}m_{\sigma_0}^2},\\
 &V_{\eta'N\rightarrow \eta N}=\frac{6gB}{\sqrt{6}m_{\sigma_8}^2}. 
\end{align}
Here, $m_{\sigma_0}$ and $m_{\sigma_8}$ are the singlet and octet scalar
meson masses, respectively.
The parameter $B$ is the coefficient of the determinant term, so it
reflects the U$_A$(1) anomaly.
The $\eta'N$ transition to the $\eta N$ channel is relatively suppressed
by the larger octet scalar-meson mass compared with $V_{\eta'N}$.

Evaluating the amplitude at the $\eta'N$ threshold, we find
that the interaction is comparably strong to the Weinberg-Tomozawa
interaction in the $\bar{K}N$ channel with $I=0$ where there exists the
$\Lambda(1405)$ as a quasi-bound state \cite{Hyodo2012}.

\section{The analysis of the $\eta'N$ two-body system}
In this section, we explain the analysis of the $\eta'N$ system and show
the result of our calculation.

We analyze the $\eta'N$ system with the analogous method to the
$\bar{K}N$ system \cite{Hyodo2012}; the scattering amplitude from the
tree-level chiral perturbation theory is used as the interaction kernel.
The divergence of the loop integral is renormalized with the natural
renormalization scheme \cite{Hyodo2008}.
This means that we eliminate the dynamics other than the $\eta'$ and
nucleon with this scheme.

The $T$ matrix obeys the scattering equation,
\begin{align}
 T_{\alpha\beta}(P)&=V_{\alpha\beta}+V_{\alpha\gamma}G_\gamma(P)
 T_{\gamma\beta}(P)
\end{align}
where the subscript $\alpha,\beta,\gamma$ are the label of the
channel and they can be $\eta'N,\eta N,$ or $\pi N$, and $P$ is the
total momentum of the two particles.
$G(P)$ is the two-body Green's function and its divergence is
renormalized with the natural renormalization scheme.
Now, we use the scattering amplitude obtained from the linear sigma
model at the $\eta'N$ threshold as the interaction kernel.
The interaction kernel is momentum independent, so the scattering
equation can be solved in the algebraic way.

In Table \ref{epN_2body_results}, we present the values obtained from
the analysis of the $T$ matrix of the $\eta'N$ channel \cite{Sakai2014}.
\begin{table}
\centering
\caption{The obtained values of the $\eta'N$ system \cite{Sakai2014}.}
\begin{tabular}{ccc}
\hline\noalign{\smallskip}
$\eta' N$ binding energy [MeV] &$\eta'N$ scattering length [fm]&$\eta'N$
 effective
 range [fm] \\
\noalign{\smallskip}\hline\noalign{\smallskip}
12.3-3.3i&-1.91+0.24i &0.24-7.6$\times 10^{-3}i$ \\
\noalign{\smallskip}\hline
\end{tabular}
\label{epN_2body_results}
\end{table}
From the search of the pole in the complex energy plane, we find a pole
corresponding to the $\eta'N$ bound state with the binding energy
$12.3-3.3i$ MeV.
The scattering length and effective range of the $\eta'N$ system are
$-1.91+0.24i$ fm and $0.24-7.6\times 10^{-3}i$ fm, respectively.
The scattering length is the repulsive sign due to the existence of the
$\eta'N$ bound state.
Compared with the scattering length suggested from the analysis of the
experimental data which are less than 1 fm \cite{Moskal2000,Moskal2014},
the obtained scattering length is large value.

\begin{figure}
 \centering
  \includegraphics[width=7cm]{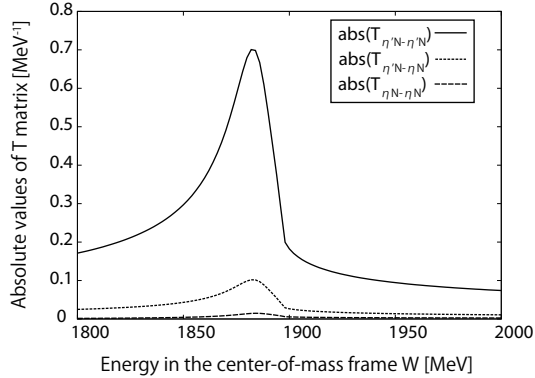}
\caption{The absolute values of the T matrices \cite{Sakai2014}.
The solid, dotted, and dashed lines represent the channel of the
 $\eta'N$ to $\eta'N$, $\eta'N$ to $\eta N$, and $\eta N$ to $\eta N$,
 respectively.
The threshold of the $\eta'N$ channel is 1896.7MeV.
}
\label{epN_abs_tmtrx}       
\end{figure}

The plot of the absolute value of the $T$ matrix is shown in
Fig.~\ref{epN_abs_tmtrx} \cite{Sakai2014}.
One can see the sub-threshold peak coming from the $\eta'N$ bound state
in Fig.~\ref{epN_abs_tmtrx}.

\section{Summary}
\label{summary}
Using the linear sigma model, we investigate the $\eta'N$ 2body
interaction with is essential for the discussion of the in-medium
$\eta'$ property.

In the linear sigma model, the $\eta'N$ interaction is generated by the
anomaly-induced scalar meson exchange, which is quite different origin
from the ordinary NG boson-nucleon interaction.
The transition to the $\eta$ mesons is relatively
suppressed by the large mass of the octet scalar meson compared with
the $\eta'N$ elastic channel.

From the analysis of the $\eta'N$ system,
we find a bound state with the binding energy 12.3 MeV and
the half width $3.3$ MeV. 
The obtained scattering length is about $-1.91+0.24i$ fm whose real part
has the repulsive sign.

\begin{acknowledgements}
S.~S. appreciates the support by the  Grant-in-Aid for JSPS Fellows
 (No.~25-1879).
This work was partially supported by Grants-in-Aid for Scientific Research
from MEXT and JSPS (No. 25400254 and No. 24540274).
\end{acknowledgements}


\begin{thebibliography}{}
%
%
\bibitem{Cohen1996}T.D.~Cohen, Phys.~Rev.~{\bf D54}, 1867 (1996).
\bibitem{Lee1996}S.H.~Lee and T.~Hatsuda, Phys.~Rev.~{\bf D54}, 1871
	(1996).
\bibitem{Evans1996}N.~Evans, S.D.H.~Hsu, and M.Schwetz, Phys.~Lett.~{\bf
	B375}, 262 (1996).
\bibitem{Jido2012}D.~Jido, H.~Nagahiro, S.~Hirenzaki, Phys.~Rev.~{\bf
	C85}, 032201 (2012).
\bibitem{Shuryak1982}E.V.~Shuryak, Nucl.~Phys.~{\bf B203}, 140 (1982).
\bibitem{Pisarski1984}R.D.~Pisarski, F.~Wilczek, Phys.~Rev.~{\bf D29},
	338 (1984).
\bibitem{Kunihiro1989}T.~Kunihiro, Phys.~Lett.~{\bf B219}, 363 (1989);
	 Phys.~Lett.~{\bf B245}, 687 (E).
\bibitem{Drukarev1991}E.G.~Drukarev and E.M.~Levin,
	Prog.~Part.Nucl.~Phys.~{\bf 27}, 77 (1991).
\bibitem{Suzuki2004}K.~Suzuki, {\it et al}., Phys~Rev.~Lett.~{\bf 92},
	72302 (2004).
\bibitem{Friedman2004}E.~Friedman, {\it et al.}, Phys.~Rev.~Lett.~{\bf
	93}, 122302 (2004).
\bibitem{Bernard1988}V.~Bernard, U.G.~Meissner, Phys.~Rev.~{\bf D38},
	1551 (1988).
\bibitem{Costa2003}P.~Costa, M.C.~Ruivo, and Y.I.~Kalinovsky,
	Phys.~Lett.~{\bf B569}, 171 (2003).
\bibitem{Nagahiro2006}H.~Nagahiro, M.~Takizawa, and S.~Hirenzaki,
	Phys.~Rev.~{\bf C74}, 045203 (2006).
\bibitem{Bass2006}S.D.~Bass and A.W.~Thomas, Phys.~Lett.~{\bf B634}, 368 (2006).
\bibitem{Sakai2013}S.~Sakai and D.~Jido, Phys.~Rev.~{\bf C89}, 041901 (2013).
\bibitem{Nagahiro2005}H.~Nagahiro and S.~Hirenzaki,
	Phys.~Rev.~Lett.~{\bf 94}, 232503 (2005).
\bibitem{Nanova2013}M.~Nanova, {\it et al}., Phys.~Lett.~{\bf B727}, 417
	(2013).
\bibitem{Moskal2000}P.~Moskal, {\it et al}., Phys.~Lett.~{\bf B482},
	356 (2000).
\bibitem{Moskal2014}P.~Moskal, {\it et al}., Phys.~Rev.~Lett.~{\bf 113},
	062004 (2014).
\bibitem{Hyodo2012}T.~Hyodo and D.~Jido, Prog.~Part.~Nucl.~Phys.~{\bf
	67}, 55 (2012).
\bibitem{Hyodo2008}T.~Hyodo, D.~Jido, and A.~Hosaka, Phys.~Rev.~{\bf
	C78}, 025203 (2008).
\bibitem{Sakai2014}S.~Sakai and D.~Jido, in preparation.
\end{thebibliography}


\end{document}